\documentclass[11pt,twoside]{article}
\usepackage{newpasp}

\usepackage{epsf}

\index{Kim, S.}
\index{Staveley-Smith, L.}
\index{Sault, R.J.}

\begin{document}

\title{Gas \& Galaxy Evolution: HI Imaging of the Large Magellanic Cloud}
\author{S. Kim}
\affil{SAO, 60 Garden St., MS-78, Cambridge, MA, 02138}
\author{L. Staveley-Smith, R.J. Sault}
\affil{ATNF, 76 Epping, CSIRO, NSW, 2121, Australia}

\begin{abstract}
We present results from the combined Australia Telescope Compact Array (ATCA) 
aperture synthesis mosaic survey of HI emission in the Large Magellanic Cloud
with 64m Parkes single dish telescope observations.
\end{abstract}




\section{Introduction}

The 21-cm line emission from the Large Magellanic Cloud (LMC) was first 
detected by Mills (1953) and a preliminary survey of its distribution was 
conducted by Kerr et al. (1954). The full HI structure of the LMC was 
first surveyed by McGee (1964) and McGee and Milton (1966) using the 
Parkes single dish radio telescope. The overall distribution of the 
neutral hydrogen gas in the LMC and its relation to the Small Magellanic
Cloud (SMC) and the Galaxy were first described in those papers. More 
recently, the details of the HI distribution and dynamics in the LMC have 
been re-examined with the Parkes telescope by Rohlfs et al. (1984) and 
Luks and Rohlfs (1992). 

Recent high resolution HI sysnthesis observations using the Australia 
Telescope Compact Array (ATCA) of the LMC (Kim et al. 1998) and the SMC 
(Staveley-Smith et al. 1997; Stanimirovic et al. 1999) have provided the 
detailed HI structure and can be used to show the correlation between the 
surface density of HI gas and the HII regions and stellar objects. 
This paper presents the combined ATCA interferometric survey of neutral 
hydrogen in the LMC with new Parkes telescope observations and provide a 
set of images sensitive to all angular spatial scales. 

\section{Observations}

The ATCA was used in the mosaic mode to survey a region 10$^{\circ}$
$\times$ 12$^{\circ}$ covering the LMC, and at an angular resolution
of $1^{\prime}$, corresponding to a spatial resolution of 15 pc (for
an assumed distance of 50 kpc). The detailed observations and data
reduction are described in Kim et al. (1998). The ATCA data for the LMC 
have been combined with Parkes single-dish data in order to correct for 
the spatial filtering properties of the interferometer. 

The Parkes observations were taken with the inner 7 beams of the Parkes 
Multibeam receiver in December 1998. The telescope has a half-power 
beamwidth of 14\farcm1. A total bandwidth of 8 MHz was used with 2048 
spectral channels in each of two orthogonal linear polarizations. 
The velocity spacing of the multibeam data is 0.82 km s$^{-1}$, but the 
final cube was Hanning-smoothed to a resolution of 1.65 km s$^{-1}$. The 
velocity range in the trimmed final cube is $-66$ to 430 km s$^{-1}$ with 
respect to the barycentric reference frame. 
The ATCA map and the Parkes data cube were compared within the region of 
visibility overlap (21 $-$ 31 m) and corrected for flux calibration factor 
$f$ = 1.3 at the effective beam size of 16.9 arcmin and then combined using 
{\sc IMMERGE} in {\sc MIRIAD}. The resolution of the combined HI image of the 
LMC is 1\farcm0 which is the same as for the ATCA interferometer map. The 
pixel size in the final maps is $40^{\prime\prime}$, corresponding to $\sim10$
pc. 

\section{Result}

We show the brightest HI emission component of the LMC at each spatial 
position from all the channel maps in the heliocentric velocity range of 
$V_{hel}$ = 190 to 387 km s$^{-1}$ from the combined ATCA interferometeric 
survey in Figure 1. The HI column density image in this velocity range is 
shown in Figure 2. A spiral structure is clearly seen within outer disk as 
it is already explained in the analysis of the ATCA map alone (Kim et al. 
1998). The supergiant shells which were previously defined in the ATCA data
(Kim et al. 1999) are prominent and shells with diameters of a few tens of 
parsecs to hundreds of parsecs are still dominant features. The prominent HI 
hole is situated in the direct vicinity of the western end of the optical bar
(Kim et al. 1999).
The large HI cloud complex south of 30 Dor appears to be the counterpart for
the huge complex of molecular clouds detected in a complete CO survey by 
Cohen et al. (1988).

\begin{figure}
\plotone{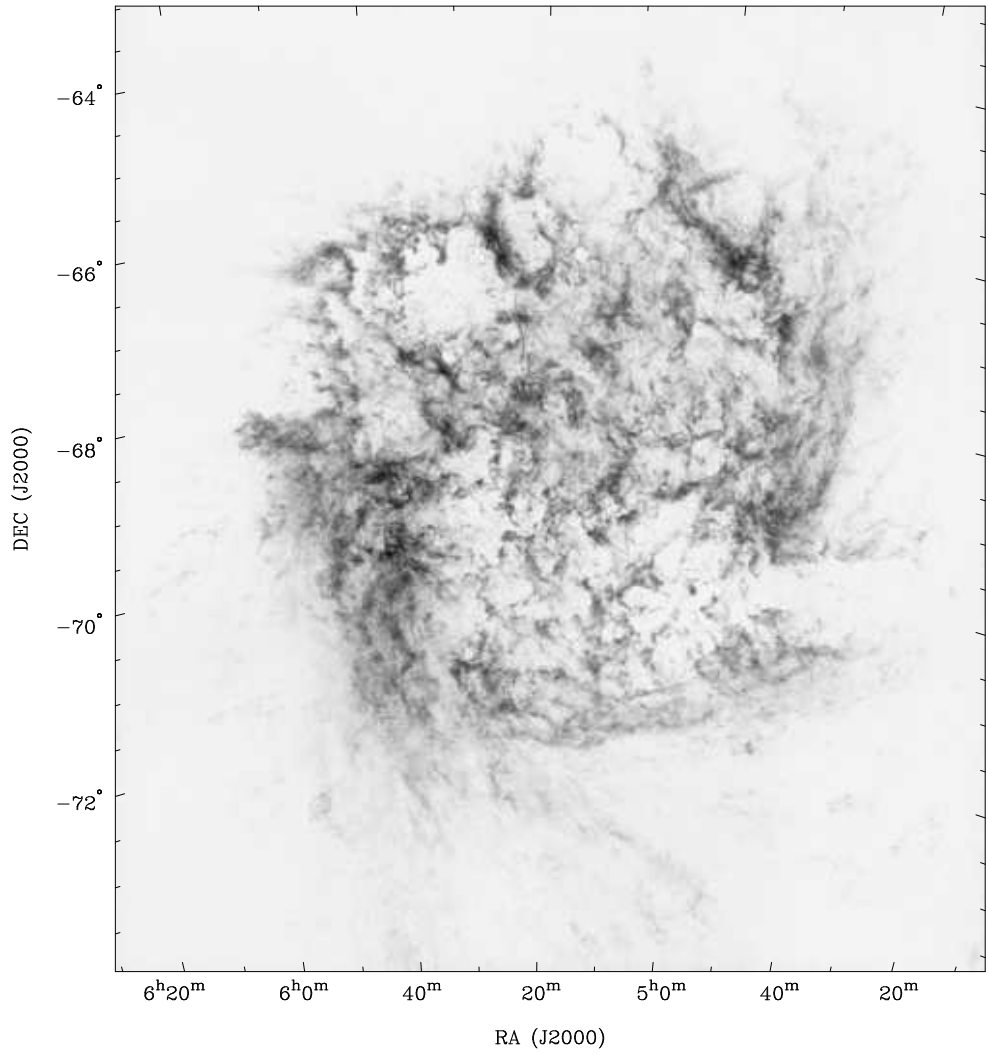}
\caption{\small The peak HI surface brightness map for the LMC. HI image 
of the LMC shows a variety of structures on different scales.}
\label{fig1}
\end{figure} 

\begin{figure}
\plotone{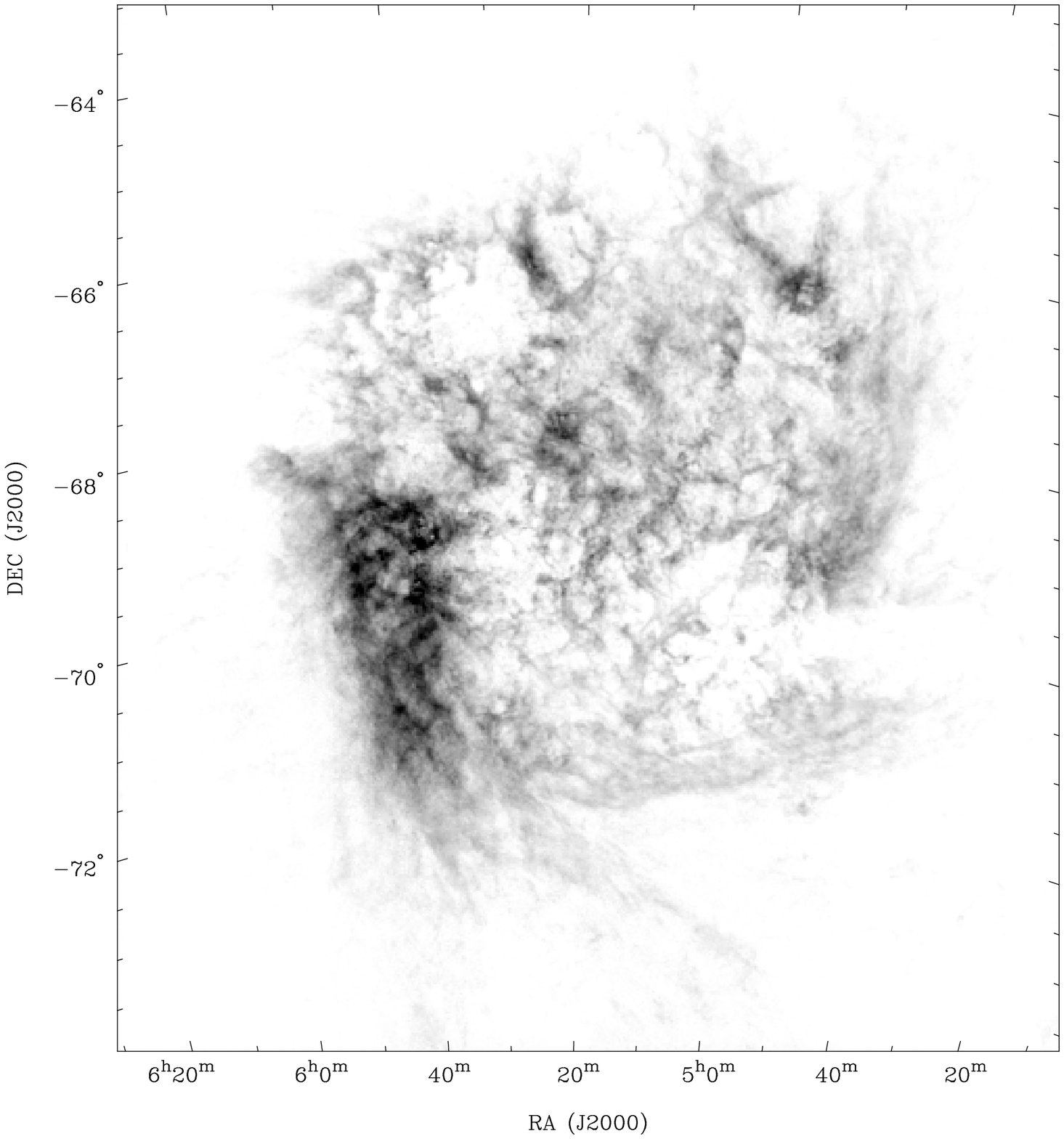}
\caption{\small HI column density image of the LMC. The grey-scale intensity
range is 0 to 5 $\times$ 10$^{21}$ H-atom cm$^{-2}$ in HI column density.} 
\label{fig2}
\end{figure}

\acknowledgments

SK thanks to Michael A. Dopita, Ken Freeman, You-Hua Chu, Mike Kesteven, 
and Dave McConnell.

\end{document}